# Towards Machine Learning-based Quantitative Hyperspectral Image Guidance for Brain Tumor Resection

David Black[1], Declan Byrne[1], Anna Walke[2], Sidong Liu[3], Antonio Di Ieva[3], Sadahiro Kaneko[4], Walter Stummer[2], Tim Salcudean[1], Eric Suero Molina[2,3*]

[1]Department of Electrical and Computer Engineering, University of British Columbia, Vancouver, Canada
[2]Department of Neurosurgery, University Hospital of Münster, Münster, Germany.
[3]Computational NeuroSurgery (CNS) Lab, Macquarie Medical School, Faculty of Medicine, Health and Human Sciences, Macquarie University, Sydney, NSW, Australia.
[4]Department of Neurosurgery, Hokkaido Medical Center, National Hospital Organization

**Corresponding author:**
PD Dr. med. Eric Suero Molina, MBA
Department of Neurosurgery
University Hospital of Münster
Albert-Schweitzer-Campus 1, A1
D-48149 Münster
eric.suero@ukmuenster.de
Tel. +49 251 83 47472
Fax +49 251 83 47479

Abstract

Complete resection of malignant gliomas is hampered by the difficulty in distinguishing tumor cells at the infiltration zone. Fluorescence guidance with 5-ALA assists in reaching this goal. Using hyperspectral imaging, previous work characterized five fluorophores' emission spectra in most human brain tumors. In this paper, the effectiveness of these five spectra was explored for different tumor and tissue classification tasks in 184 patients (891 hyperspectral measurements) harboring low- (n=30) and high-grade gliomas (n=115), non-glial primary brain tumors (n=19), radiation necrosis (n=2), miscellaneous (n=10) and metastases (n=8). Four machine learning models were trained to classify tumor type, grade, glioma margins and IDH mutation. Using random forests and multi-layer perceptrons, the classifiers achieved average test accuracies of 84-87%, 96.1%, 86%, and 93% respectively. All five fluorophore abundances varied between tumor margin types and tumor grades ($p < 0.01$). For tissue type, at least four of the five fluorophore abundances were found to be significantly different ($p < 0.01$) between all classes. These results demonstrate the fluorophores' differing abundances in different tissue classes, as well as the value of the five fluorophores as potential optical biomarkers, opening new opportunities for intraoperative classification systems in fluorescence-guided neurosurgery.

Introduction

Surgical resection of malignant glioma is complex, and recurrences are the rule rather than the exception, leading to patients' poor prognosis[1]. This is partly due to poorly differentiated tumor tissue, especially at infiltrating margins, closely resembling healthy tissue during surgery[2]. To address this problem, 5-aminolevulinic acid (5-ALA)-induced protoporphyrin IX (PpIX) fluorescence guidance has been established as a surgical adjunct in neurosurgery. The complete resection rate for contrast-enhancing tumors increased, in the initial approval study from 2006, from 36% operated under white light only to 65% for resections with fluorescence guidance[2]. Improvements in neuroimaging and different surgical adjuncts, such as intraoperative ultrasound, pre- and intra-operative brain mapping, and monitoring techniques, amongst others, currently allow for over 95% complete resection rates, whenever feasible[3]. Consequently, PpIX fluorescence is widely used in the resection of high-grade gliomas. Furthermore, it is the subject of research in low-grade gliomas[4–7], meningiomas[8–11], and other brain tumors[12], as well as in oral cancer[13,14], bladder cancer[15,16], and skin cancer[17]. Additionally, 5-ALA is used for photodynamic therapy[18], to treat skin and other brain malignancies[19].

Fluorescence is achieved by administering 5-ALA orally prior to surgery[1], which is metabolized to PpIX, a precursor in heme biosynthesis. The mechanisms leading to selective accumulation of PpIX in glioma cells[20] are not entirely understood[1,11,21]. Several explanations have been proposed, including disruption of the blood-brain barrier, as commonly observed in high-grade gliomas (HGG), which is otherwise non-permeable to 5-ALA[1,7,22]. Reduced activity of ferrochelatase, which would otherwise metabolize PpIX[15,21,23], and changes in the tumor microenvironment affecting 5-ALA uptake and PpIX efflux[24–26] may play a part as well. However, PpIX is present in increased concentrations 7-8 hours after 5-ALA administration in HGG[27]. PpIX fluoresces after illumination with intense 405 nm (blue) light. Absorption of such a photon lifts the molecule to an excited state from which it decreases slightly through vibrational relaxation before returning to its ground state and emitting a second photon[28]. Due to the decrease in energy before the second transition, the emitted photon has a longer wavelength of around 634 nm (red). This energy change is called the Stokes shift and allows the red fluorescing areas to be differentiated from the otherwise blue-reflecting tissue[29].

With optical highpass filters such as the BLUE400 and BLUE400 AR systems (Carl Zeiss Meditec AG, Oberkochen, Germany) that block the relatively intense reflected blue light but transmit the red fluorescence, surgeons are able to differentiate 5-ALA positive tumors from healthy tissue[30,31]. This has led to widespread adoption. However, the proven increased resection rates are only for malignant glioma. Unfortunately, many lower-grade gliomas and even some with high-grade regions exhibit low PpIX accumulation and do not visibly fluoresce. Furthermore, autofluorescence makes it impossible to discern the diminished PpIX fluorescence even when imaging sensitivity is increased because it shares the same spectral range as the PpIX emission[32–35].

Hence, quantitative spectroscopic systems have been developed that measure the emission spectra and separate PpIX from autofluorescence through a priori knowledge of the fluorophores present and their individual emission spectra[35–37]. These devices have been used primarily in research and consist of either point probes[5,8,37–39] or wide-field hyperspectral[12,40,41] devices. In addition to potential future intraoperative use to help distinguish tumors, spectroscopy can be used to answer many questions about different diseases[23,36], how best to treat them[27,40,41], and how to improve imaging systems for future intra-operative integration[12,42].

One potential use of such a system is to distinguish between different tissue types. Several classifications are commonly applied to brain tumors, including the tumor type (e.g., glioma, meningioma, medulloblastoma, and others[43]). Additionally, the differentiation of tumor from non-tumor or necrotic tissue (e.g., radiation necrosis, which can radiologically mimic tumor progression) is relevant in neuro-oncology and neurosurgery. As different tissue types differ substantially in their behavior and prognosis, their correct identification is paramount. Several studies have classified brain tumor type in MR images[44,45], but the same has not been attempted with fluorescence.

Furthermore, gliomas are classified according to their histological, genetic features, and biological behavior[46]. They are categorized as grade I-IV by the World Health Organization (WHO) classification system, where I and II are considered low grade, and III and IV are high grade[47], indicating a bad prognosis[48]. Molecular parameters, i.e., isocitrate dehydrogenase (IDH)-mutation, O6-methylguanine-DNA-methyltransferase (MGMT), among others, assist in further subclassification of these tumors[49]. We aim to predict tumor molecular characteristics, such as IDH mutations or genetic aberrations related to malignancy by correlation with the spectral signature of tumors. Such knowledge would directly result in changes to surgical strategy. IDH mutations cause a shift in enzymatic activity, converting α-ketoglutarate to 2-hydroxyglutarate and inhibiting α-KG-dependent enzymes. This leads to metabolic reprogramming, hinders cell differentiation, and initiates tumorigenesis[50]. Small molecule inhibitors can reverse this process, making knowledge of IDH mutation status crucial for guiding their targeted use in treatment. Knowing a tumor's grade or molecular characteristics could also help make decisions during and after surgery. The tumor types are shown in Fig. 1. Some studies have performed automatic classification of tumor grade using convolutional neural networks (CNN) on magnetic resonance images[51] and digitized histopathology slides[52], but not through fluorescence.

Additionally, one of the primary difficulties of resection of grade II, III, and IV gliomas is the infiltrative nature of the tumors. Around the solid tumor portion, there is a region of infiltrative margin characterized by decreased tumor cellularity transitioning to healthy tissue[53]. To delineate the margins, guidance from pre-operative MRI images is commonly used in intraoperative neuronavigation systems[54], but its accuracy suffers from factors such as brain shift[55]. Differentiating more accurately between margin regions would significantly enhance the effectiveness of surgical resection of glioma and patients' outcomes[56,57]. Leclerc et al. presented the first work applying machine learning (ML) to neurosurgical fluorescence spectroscopy, classifying different tissues based on principal component analysis (PCA) of the fluorescence spectra[58], achieving 77% accuracy on 50 samples.

Although previous work has correlated PpIX fluorescence with WHO grade and tumor margins[59], it is not well understood how different tumor types affect the abundances of the different fluorophores[60]. Through recent characterization of their basis spectra[35], we can now precisely analyse the abundances of the major fluorophores in hyperspectral fluorescence images. Hence, it is possible to study how the two photo-states of PpIX and the autofluorescence from flavins (i.e., flavin adenine dinucleotide), NADH, and lipofuscin are affected by the tissue type. This paper performs four classification tasks using fluorescence spectroscopy, aiming to use the abundances of the five fluorophores to differentiate between (1) tumor and tissue types (e.g., glioblastoma, meningioma, etc.), (2) WHO grades, (3) between solid tumor, infiltrating zone, and reactively altered non-tumor brain tissue and (4) between IDH-mutant and IDH-wildtype glioma.

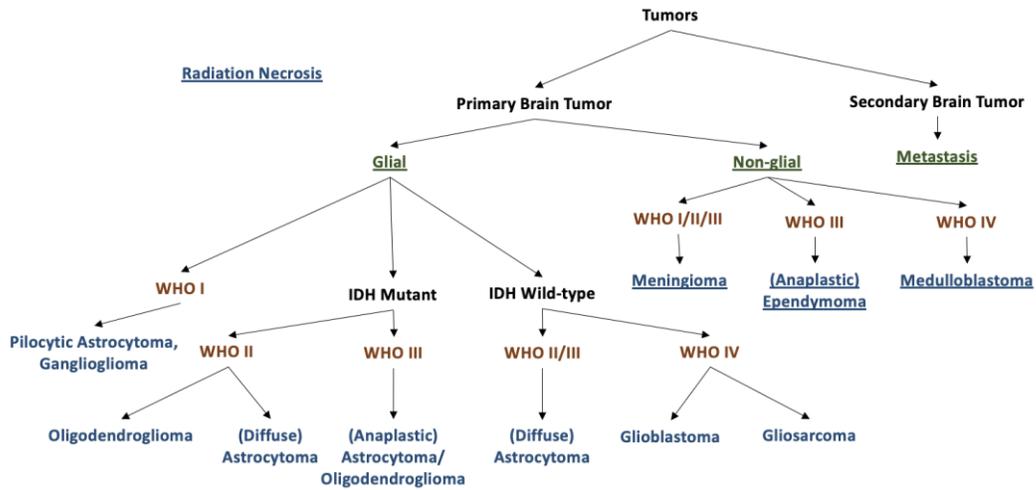

**Figure 1.** Taxonomy of tissue types considered in this study. As many tumors were measured before 2021, this partly uses the 2016 classification. WHO grade (orange) is shown in the WHO Grade Section of the Results. Tissue type classification (blue) and higher-level group classification (underlined) are described in Tissue Type Section of the Results.

## Methods

### Device and Dataset

A hyperspectral imaging device was used, as previously described[27,41,59] and shown in Fig. 9. Patients received 5-ALA four hours prior to anesthesia induction orally at a dose of 20mg/k.g. b.w. All experiments complied with institutional guidelines and were approved by the ethical committee of the University of Münster. Tissue samples were resected during surgery and measured directly *ex vivo* on a petri dish. The sample was first illuminated with blue light (405 nm), then with no light for recording background noise, and finally with broadband white light. During each illumination phase, the emitted light from fluorescence and the reflected light were gathered in the objective lens and routed to a scientific metal oxide semiconductor (sCMOS) camera through a series of optical filters. The first filters removed the intense blue reflected light. Next, a liquid crystal tunable bandpass filter (LCTF) transmitted a narrow spectral band of the emitted light to the sCMOS. The LCTF was swept across the visible range (420-730 nm) in 3-5 nm increments, during each of which a grayscale image was captured by the sCMOS. In this way, a hyperspectral data cube was generated, containing all the spectral and spatial information. Any pixel can be selected, and an emission spectrum extracted for that pixel from the cube. During blue-light illumination, the fluorescence spectra were captured. The white-light illuminated data cube was used for dual-band normalization[42], and the no-light spectra were used to remove dark noise of the camera sensor from the images.

This device was used for *ex vivo* analysis of tumor biopsies at the University Hospital of Münster, Germany. The resulting data set was used in this study to explore the effect of different classifications of tumors on the presence of the five main fluorophores. In particular, we analyzed tumor type, WHO grade, and whether the sample is from solid tumor, infiltrating zone, or non-tumor. The classes and number of samples are shown below and were chosen based on the availability of patient data. Many samples were measured before the updated WHO classification (CNS5, 2021)[47] and thus followed the previous 2016 WHO classification system[49].

*Tissue Type Classification:* (n = 632 biopsies)
Pilocytic astrocytoma (PA; n=5), diffuse astrocytoma (DA; n=57), anaplastic astrocytoma (AA; n=51), glioblastoma (GB; n=410), grade II oligodendroglioma (OD; n=24), ganglioglioma (GG, n=4), medulloblastoma (MB; n=6), anaplastic ependymoma (AE; n=8), anaplastic oligodendroglioma (AO; n=4), meningioma (MN; n=37), metastasis (MT; n=6), radiation necrosis (RN; n=20).

*Margin Classification:* (n = 288 biopsies)
Solid tumor (ST; n=131), infiltrating zone (IZ; n=57), reactively altered brain tissue (RABT; n=100).

*WHO Grade Classification:* (n = 571 biopsies)
Grades I (n=9), II (n=84), III (n=57), IV (n=421).

*IDH Classification:* (n= 411 biopsies)
IDH-mutant (n=126), IDH-wildtype (n=285)

The biopsies varied in size and shape, but approximately 100-1000 spectra were extracted from each. Regions of 10x10 pixels were averaged to produce one spectrum for noise reduction. For a given classification task, spectra were sampled randomly from the tumor portions of the biopsies. To avoid inadvertently sampling from the background glass slide, the tumor was first segmented automatically in MATLAB using the 634 nm fluorescence image. The result is shown in step 2 of Fig. 9; resulting masks were also checked manually.

The raw fluorescence spectra were corrected, normalized, and unmixed as described previously[12,35]. Each fluorescence spectrum is assumed to consist of a linear combination of five fluorophores whose spectra are known a priori. These basis spectra are the two photo-states of PpIX (denoted PpIX634 and PpIX620)[36,37], NADH, lipofuscin, and flavins[35]. The unmixing calculates the abundances of the five spectra by minimizing the squared error between the measured spectrum and a linear combination of the five spectra. If the basis spectra are combined as columns in a 310x5 matrix $B$, the measured spectrum is the 310x1 vector $y$, and the 5x1 vector of abundances is $c$, then we find $c$ using $c = \underset{c}{\mathrm{argmin}}\left\{\frac{1}{2}\|y - Bc\|^2\right\}$, using non-negative least squares. This is illustrated in steps 3-5 of Fig. 9.

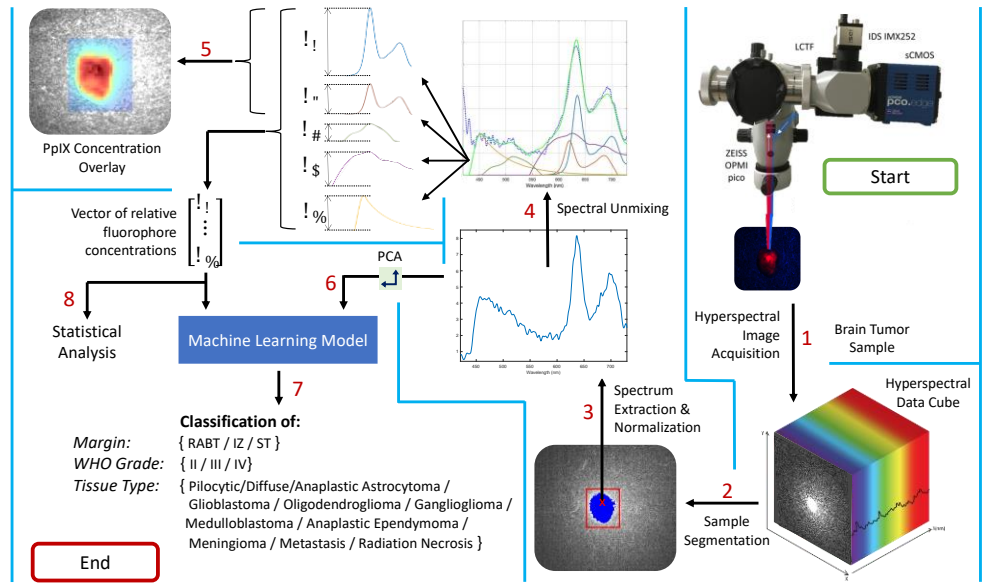

**Figure 9.** Overview of the device and method.

*Machine Learning Approach*

With the processed data, we tried several ML classification models using Python and experimented with different hyperparameters. The different types are listed in Table 3.

| Pixels per Sample | [1, 2, 3] | | |
|---|---|---|---|
| Samples per Class | Tumour Type | Margin | WHO Grade |
| | [300, 500, 800] | [1000, 3000, 5000] | [500, 1500, 3000] |
| Random Forests[61] | Number of trees: [50, 75, 100, 125, 150] | | |
| | Splitting Criteria: [Gini, Entropy, Log loss] | | |
| | Minimum samples to split: [2, 3, 4] | | |
| | Maximum features per tree: [sqrt, log2, None] | | |
| KNNs[62] | Number of neighbors: [3, 5, 7, 9] | | |
| | Weights: [Uniform, Distance] | | |
| | p: [1, 2] | | |
| SVM[63] | Kernel: [RBF, Linear, Polynomial, Sigmoid] | | |
| MLP[64] | Number of hidden layers: [1,2,3] | | |
| | Number of neurons: [25, 50, 100, 150] | | |
| | Activation: ReLU (Rectified linear unit) | | |
| | Solver: [Adam, Limited-memory BFGS (LBFGS)] | | |
| | Nesterov Momentum: 0.9 | | |
| AdaBoost[65] | Number of estimators: 50 | | |

|  | Learning rate: 1.0<br>Algorithm: SAMME.R |

**Table 3.** Explored classifier algorithms and hyperparameters. Multi-layer perceptron (MLP), support vector machine (SVM), and k-Nearest Neighbor (KNN) classifiers are described in the papers cited in the table.

For a given biopsy, instead of using the abundances inferred from only one pixel, it could be more informative to use those of two or more nearby pixels. This is attempted in the "pixels per sample" hyperparameter. Further, to avoid bias, the same number of samples of each class were used to train the classifiers. Since some classes had fewer samples than others, we varied the number of samples per class. Lower numbers of samples per class allowed more classes to be included in the classification. However, larger numbers of samples per class allowed for more effective training of the classifiers, though on fewer classes. Classes with few biopsies, including anaplastic oligodendroglioma and ganglioglioma, had to be excluded for some tests.

The dataset was split 80/20 into a training and testing set. Hyperparameter and model tuning was carried out using 5-fold cross-validation on the training set to mitigate optimization bias. In total, 195 models were trained and compared for each classification task. For each of the four classification tasks, nine different datasets were used with different pixels per sample and samples per class. Finally, the best model was selected for each category and evaluated with the test set. To evaluate the multi-class classifiers, receiver operating characteristic curves (ROC) were determined, and the area under the curve (AUC) was computed. Accuracy was calculated as the number of correctly classified samples divided by the total number of samples. Confusion matrices were used to determine which classes performed better than others.

We also explored whether the calculated abundances were the most informative space to characterize the biopsies. Instead of relying on the five basis spectra, we repeated the above experiments with the mathematically optimal projection of the spectra onto a 5-dimensional hyperplane using principal component analysis (PCA). We used the same number of pixels per sample, samples per class, and all other hyperparameters as above. We also visualized the different classes using only the first two or three principal components in a scatter plot. PCA finds a set of orthogonal axes, which maximizes the variance. Depending on assumptions about the distributions of the signal and noise, this also maximizes the mutual information between the real signal and the dimensionally-reduced output[66]. Nonetheless, some information was lost. To quantify how well PCA represents the original data, we utilized the "variance explained" parameter. Given a data matrix of fluorescence spectra, a covariance matrix whose diagonal elements (the variances) sum to the overall variability can be computed. When performing n-component PCA, only the n principal components corresponding with the n largest eigenvalues were kept. All the eigenvalues sum to the overall variability, so by choosing only the n-largest, we lost some of the information contained in the data. In particular, the ratio of the sum of the n selected eigenvalues over the total variability was called the "variance explained" and effectively yielded the percent of total information represented by the chosen principal components.

Finally, statistical analyses were performed to determine correlations and statistical significance. This analysis was performed in MATLAB using the two-sample Kolmogorov-Smirnov test. The entire process, from imaging to data processing and extraction to analysis, is shown in Fig. 9.

## Results

As explained in detail in the Methods section, four classification tasks were explored. Tumor types belonged to the following classes: Pilocytic astrocytoma (PA), diffuse astrocytoma (DA), anaplastic astrocytoma (AA), glioblastoma (GB), grade II oligodendroglioma (OD), ganglioglioma (GG), medulloblastoma (MB), anaplastic ependymoma (AE), anaplastic oligodendroglioma (AO), meningioma (MN), metastasis (MT), and radiation necrosis (RN). Tumor margin classification was performed on three classes: solid tumor (ST), infiltrating zone (IZ), and reactively altered brain tissue (RABT). IDH status (mutated and wildtype) where assessed. Finally, WHO grades II, III, and IV were considered.

*Visualization*

Since the five known fluorophore abundances create a 5-dimensional space, it is impossible to visualize the data and distinguish patterns manually. Studying scatter plots of two fluorophores at a time would be feasible, but this would be a reductionistic approach drawing a very limited picture. Instead, we used PCA for dimensionality reduction, although this results in some information loss. The exact amount of information lost for each classification is shown in Table 1. However, PCA fails if the data forms a non-linear manifold. One alternative which captures this more effectively is t-distributed stochastic neighbor embedding (t-SNE). This was also used to better visualize the data. Some of the more informative plots are shown in Fig. 2.

| PCA Dimension | Tumor Type | Margin | WHO Grade |
| --- | --- | --- | --- |
| 2 | 79% | 79% | 85% |
| 3 | 87% | 89% | 93% |
| 4 | 94% | 95% | 95% |
| 5 | 97% | 99% | 96% |

**Table 1:** Variance explained from 2, 3, 4, and 5-dimensional PCA. A value of 100% means that all variance is accounted for. With five values, almost no information is lost, indicating that the 5 primary fluorophores most likely cause the measured emissions.

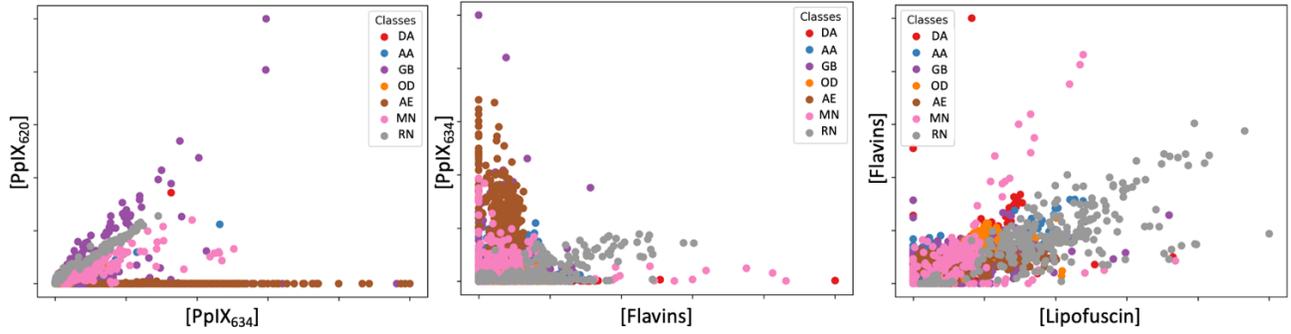

**Figure 2a.** Fluorophore abundances from different tumor types. The level of differentiation is similar to the one depicted in Fig. 2b, in which PCA was used. This suggests that the fluorophore abundances are already close to the most informative way of looking at the data.

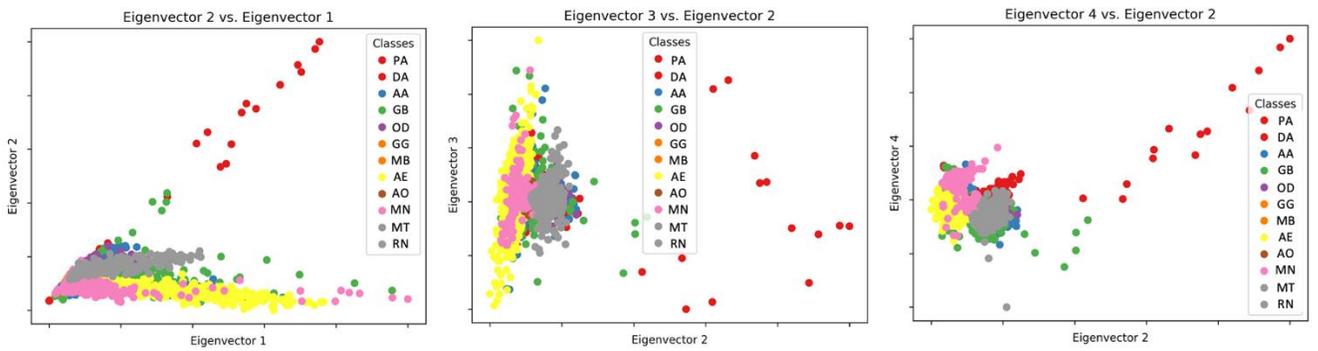

**Figure 2b.** 2-component PCA for tumor type, including more of the classes. There are some meaningful clusters even with just two components, primarily consisting of AE in yellow, MN in pink, and RN in gray. Glioblastomas are very poorly differentiated.

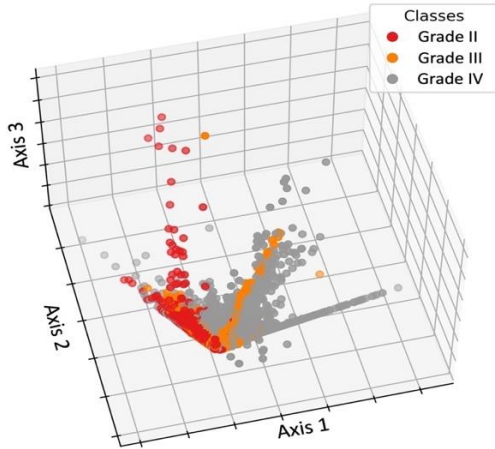 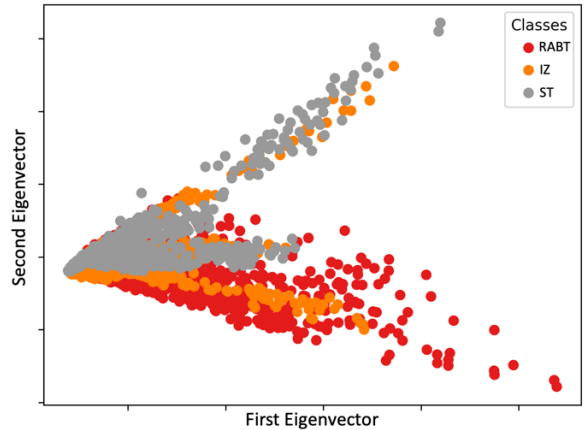

**Figure 2d.** 3-component PCA on WHO grade. It is possible to discern how a classifier may be able to distinguish between classes.

**Figure 2e.** 2-component PCA on tumor margin. This shows that IZ may be easily confused for ST or RABT, though considering the remaining three dimensions may alleviate the issue.

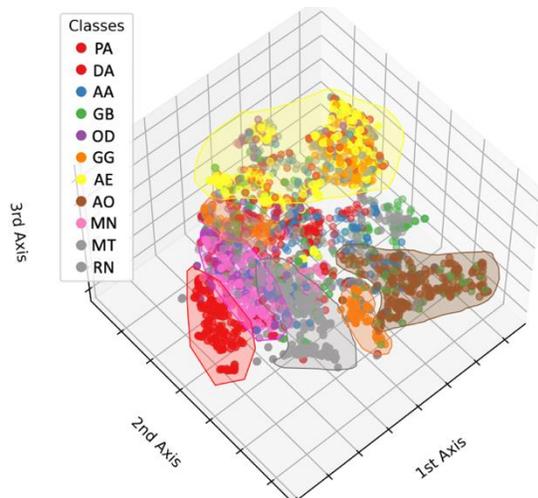

**Figure 2f.** 3-component T-SNE plot of tumor type. Here we see distinctions between some groups, though others, including glioblastoma and astrocytoma, remain very mixed.

**Figure 2.** Visualizations of tumor data for different classifications showing some promise for the feasibility and potential good performance of machine learning classifiers. Subplot d shows WHO grade, and e shows tumor margin, while the others show tumor type. WHO grade I is not considered due to its small sample size. Some tumor types are excluded due to relatively small sample size and unbalanced datasets.

Though it is difficult to draw conclusions from the reduced-dimension visualizations in Fig. 2, it shows that some classes can be distinguished from others with good confidence. Furthermore, Fig. 2a and 2b show that the fluorophores and PCA are similar, which validates the effectiveness of the spectral unmixing and the accuracy of the a priori basis spectra. Table 1 additionally demonstrates that almost all the spectral information is stored in five dimensions, which lends further credence to the five basis spectra[35]. Evidently, they are close to the mathematically optimal dimensionally reduced representation of the measured spectra.

*Tissue Type Classification*
Since the tissue type data is categorical, it is not possible to determine linear correlations. However, pairwise Kolmogorov-Smirnov tests were used to test the null hypothesis for each fluorophore: abundances from each pair of classes come from the same distribution. For each class, all available spectra were used in the significance tests, which ranged from 200 to over 5000, depending upon the class. The chosen test (kstest2.m in MATLAB) does not require equal sample sizes. All class pairs were significantly different ($p < 0.05$) across all the fluorophores except those listed in Table 2. Thus, even if a pair of classes did not differ significantly in one fluorophore, it did in the remaining four.

Both multi-layer perceptrons (MLPs) and random forests performed well for classifying the tumor type. None of the other classifiers were effective. Performance was better with more pixels per class, partly because the models were trained on more data and partly because they had to distinguish between fewer classes. Therefore, we present the best results overall and the best results from training with all the classes.

| Fluorophore | Pairs with p > 0.05 | *p*-Values |
|---|---|---|
| NADH | Radiation Necrosis | 0.11 |
| Lipofuscin | WHO grade II and III | 0.08 |
|  | Meningioma and Medulloblastoma | 0.09 |
| Flavin | Oligodendroglioma and Diffuse Astrocytoma | 0.24 |
|  | Solid Tumor and Infiltrative Zone | 0.06 |

**Table 2.** Two-sample Kolmogorov-Smirnov test for significant differences between tissue types for each fluorophore. All pairs of classes not listed in the table varied significantly (p < 0.05) for each fluorophore. PpIX abundance differed between all classes.

The best-performing classifier, including most classes, was a random forest model (150 trees, 500 samples per class, 3 pixels per sample, sqrt(n_features) features per tree, cross entropy splitting criterion). The average accuracy was 83.6%, with an AUC of 0.98 in classifying between all classes. Fig. 3 shows this classifier's confusion matrix and multiclass ROC.

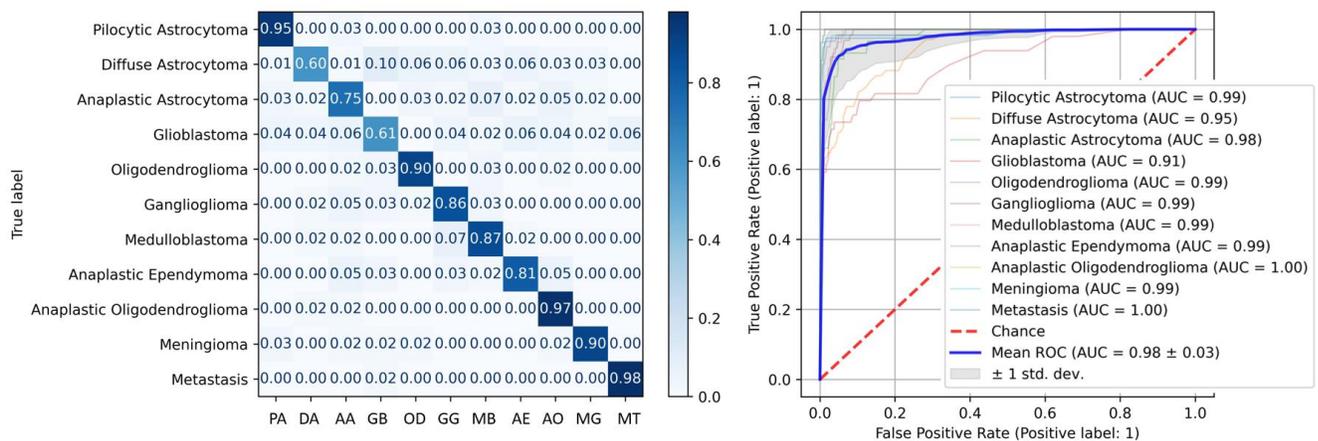

**Figure 3.** Confusion matrix and ROC for best-performing random forest classifier.

This shows very good performance for some classes and relatively poor for others. In particular, GB, DA, and AA have low classification accuracy, and GB is frequently mistaken for AE. However, AE is rare and generally easy to distinguish visually from healthy tissue, so fluorescence is not commonly used. Hence, despite its good classification accuracy, AE was removed. Additionally, this data used the 2016 WHO classification because it was partly collected before 2021. After the 2021 WHO classification, however, IDH-wildtype anaplastic astrocytomas are considered GB. Thus, these samples (n = 33) were re-labeled as GB, and the models were re-trained.

With these changes, the best-performing model used the same algorithm and hyperparameters as above but gave 87.3% accuracy and an average AUC of 0.98, as shown in Fig. 4.

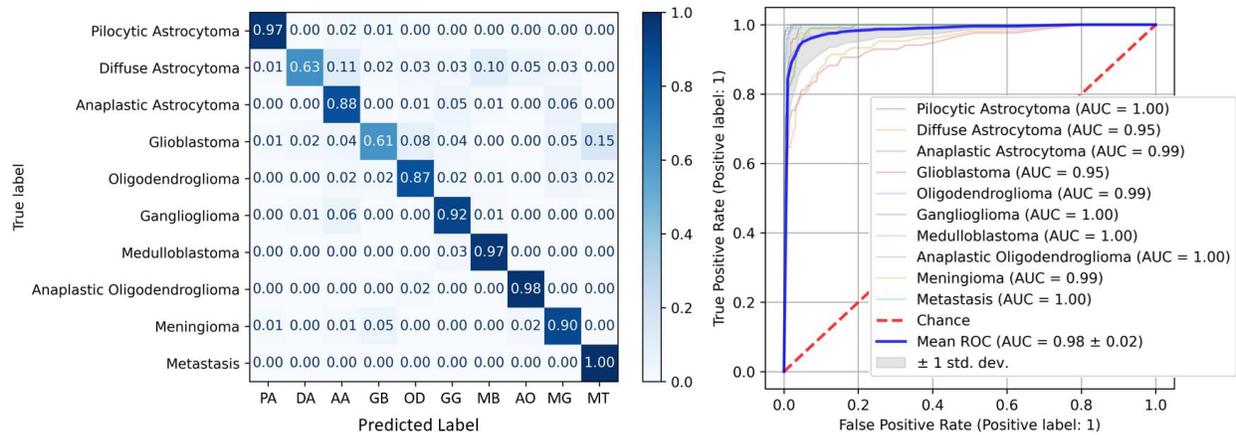

**Figure 4.** Confusion matrix and ROC for best performing random forest classifier of tumor types with AE excluded and IDH-wildtype AA relabeled as GB.

We also considered classifying between higher-level groupings of tumor types (i.e., higher in the hierarchy of Fig. 1). The classes were glioma, meningioma, medulloblastoma, and radiation necrosis; ependymoma and metastasis were excluded for small sample size. The best-performing model for this task was an MLP (150 neurons per layer, 3 hidden layers, 500 samples per class, 2 pixels per sample, Adam solver). The model achieved 89.5% accuracy and an AUC of 0.97, as shown in Fig. 5. The glioma class is very broad, which likely leads to slightly lower accuracy. Generally, however, the performance is good. The accuracy is boosted to 90.67% by using 800 samples per class, but this necessitates leaving out radiation necrosis.

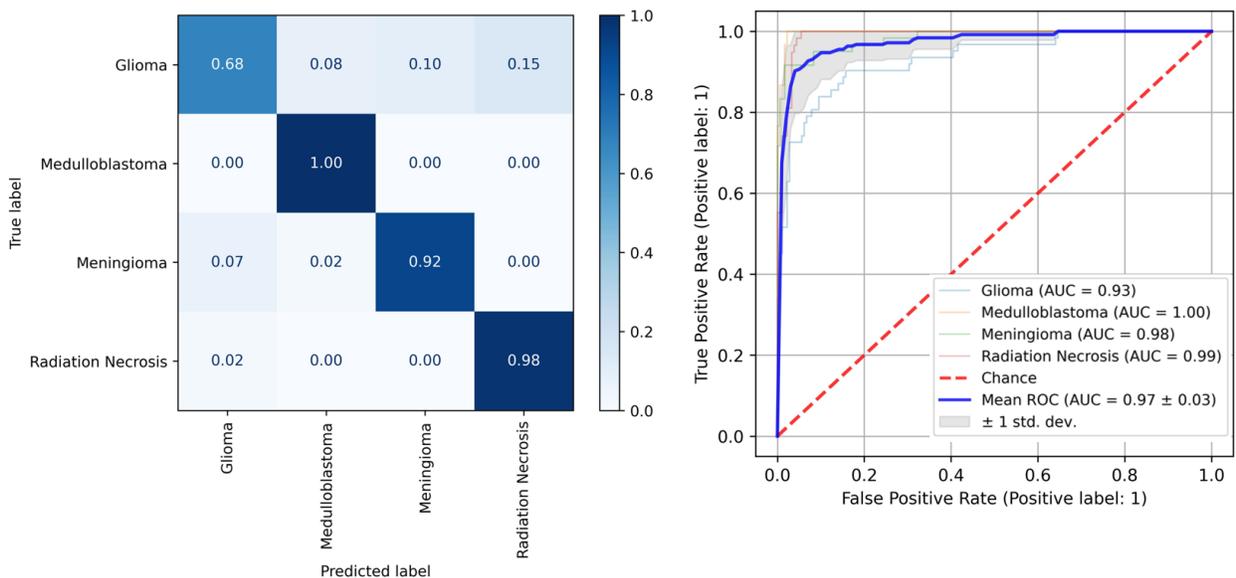

**Figure 5.** Confusion matrix and ROC for best-performing classifier of tissue type groups.

Therefore, tumor type does indeed have a strong effect on the five fluorophores: strong enough that it is possible to distinguish between eleven types of tumors and other tissue types with a relatively high degree of accuracy. Some gliomas, specifically GB, DA, and AA, were difficult to distinguish. Meningioma was also sometimes misclassified as medulloblastoma. On the other hand, many tissue types, including ganglioglioma, radiation necrosis, and to a slightly lesser degree, medulloblastoma, could be classified with almost perfect accuracy.

Using PCA instead of fluorophore abundances to perform this analysis did not improve results. This is likely due to the fact that PCA does not seem to provide more information than the fluorophores.

*Margin Classification in Gliomas*

The margin classification would be practically useful in an intraoperative setting. It could delineate solid tumor from infiltrating cells and non-tumor tissue. However, the distinction is relatively subjective (see Discussion section), and as shown in Fig. 2e, there may be substantial overlap between classes. Again, however, all fluorophore abundances are significantly different between every category ($p < 0.01$).

Consequently, a multilayer perceptron (5000 samples per class, 3 pixels per sample, 100 neurons per hidden layer, 5 hidden layers, Adam solver) distinguishes with 85.7% accuracy and a mean AUC of 0.95, as shown in Fig. 6.

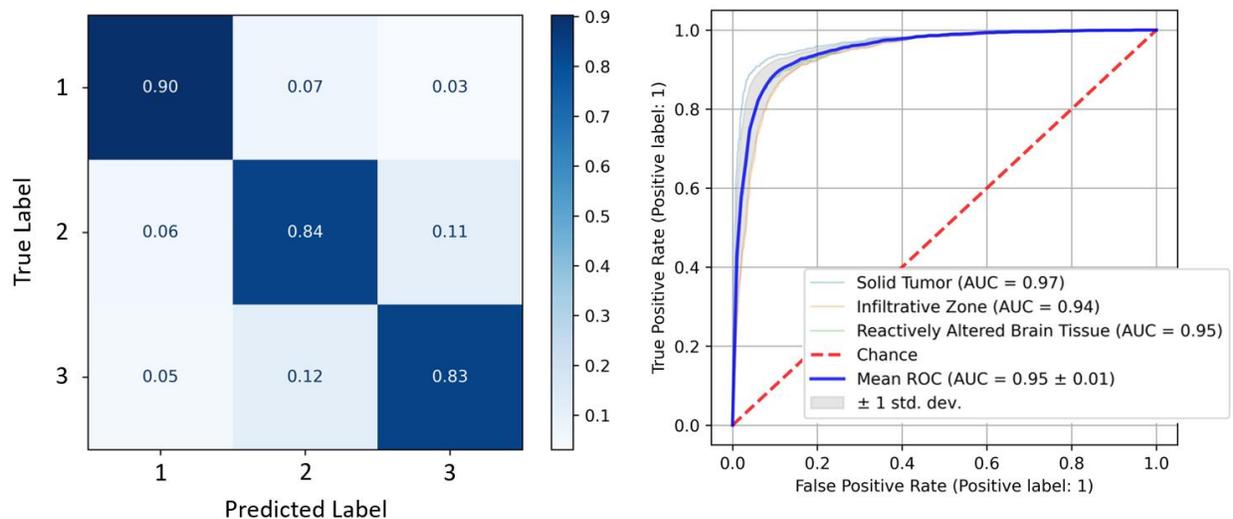

**Figure 6.** Confusion matrix and ROC for best performing MLP classifier of tumor margins, the classes are [1: Solid tumor, 2: Infiltrating zone, 3: Reactively altered brain tissue].

It is reasonable that the infiltrating zone is the most commonly mistaken for solid tumor and reactively altered brain tumor whereas the latter two are infrequently confused.

*WHO Grade Classification*

Furthermore, even though the new WHO classification is not entirely grade-based, knowing what WHO-grade certain tumors or tissue regions have during an operation would be valuable and further reinforce the malignant character of the tissue. In practice, many low-grade tumors have an anaplastic focus which fluoresces while the surrounding tumor does not. All differences in fluorophore abundances between WHO grades were found to be statistically significant ($p <$

0.01). For this task, a multilayer perceptron classifier performed best (3000 samples per class, 3 pixels per sample, 150 neurons per layer, 5 hidden layers, Adam solver), with 96.1% accuracy and a testing AUC of 0.99, as shown in Fig. 7. While WHO grades II and III were classified with good accuracy, grade III tumors were more frequently confused for grade IV.

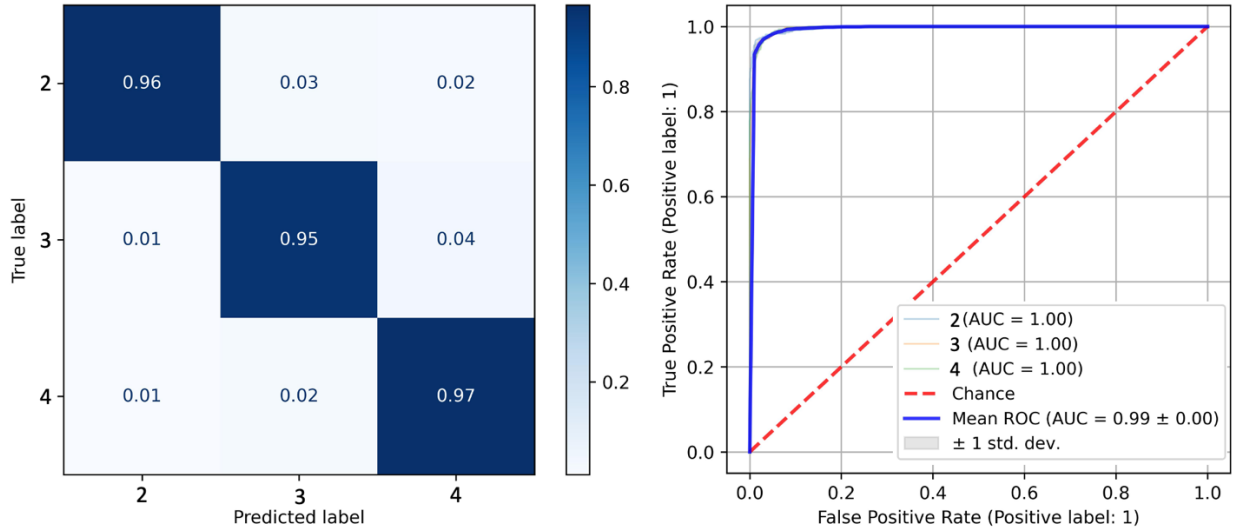

**Figure 7.** Confusion matrix and ROC for best performing random forest classifier of WHO grades, the classes are [2: Grade II, 3: Grade III, 4: Grade IV].

*IDH Mutation*

Finally, another relevant classification task is the IDH mutation. The best-performing classifier for IDH mutation was a random forest model (150 trees, all available samples per class, 2 pixels per sample, sqrt(n_features) features per tree, Shannon entropy splitting criterion). The average accuracy was 93% in classifying between IDH-mutant and IDH-wildtype tumors. Fig. 8 shows this classifier's confusion matrix and multiclass ROC.

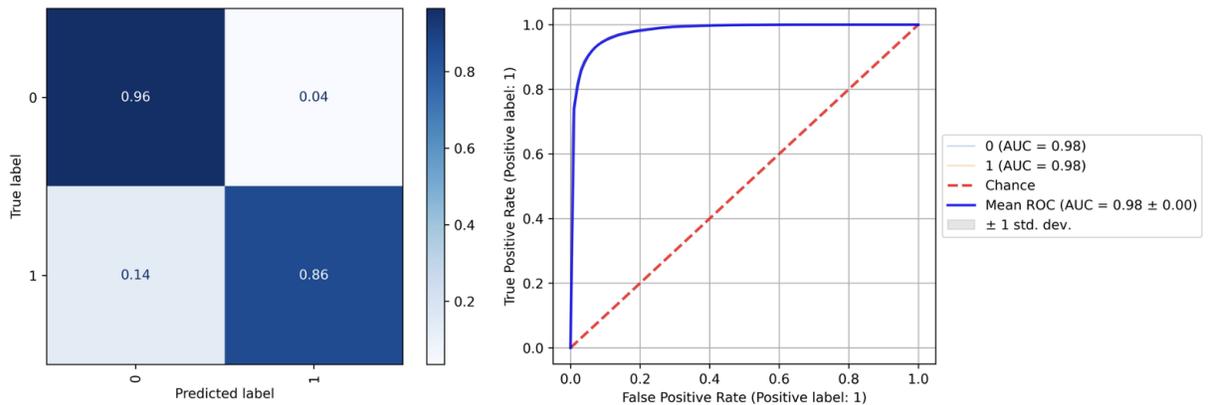

Figure 8 – Confusion matrix and ROC for best performing random forest classifier of IDH mutation (1 = mutant, 0 = wildtype). The classifier could predict IDH mutation with an accuracy of 93%.

Discussion

In this paper, four ML-based classifiers were developed for tissue types, tumor margins, and glioma WHO grades. Each classifier achieved high accuracy. The margin classification accuracy of 85.7% outperforms that of Leclerc et al.[58] (77% on 50 samples), likely due to using much more data. Our classifier also used fluorophore abundances instead of PCA. Data from 891 hyperspectral measurements of 184 patients was utilized, corresponding to up to 15000 spectra for a given test. Furthermore, the tissue type and WHO grade classifications performed well. The classifiers achieved an average test accuracy of 87.3% when classifying between tissue types and 89.5% when classifying between higher-level groupings of tissue type.

A possible limitation of the accuracies presented in this study is that some of the training data might be noisy if the automatic tumor segmentation did not work well. This is especially relevant for very small, oddly shaped biopsies or ones with a large splash of fluid next to the tumor, which tended to show up very brightly in the fluorescence images. The masks were checked manually, but some inclusion of background was inevitable. Current work is exploring the use of convolutional neural networks (CNNs) to improve the labeling process. Another source of noise in the labels is described in the Methods and Results sections, as inaccurate labeling or uncertainty in the labels could decrease training effectiveness. For example, the delineation between the margin classes is not completely straightforward, and some tumor types are no longer distinguished since the 2021 WHO classification. This was seen between Figs. 4 and 5. Future work should be carried out to replicate the results in different classes, as defined in the 2021 WHO classification system.

The results could likely be improved with more data on the less frequent tumor types (e.g., APXA, anaplastic oligodendroglioma, ganglioglioma). For the MLP and random forest models, there was a strong trend of increasing accuracy as we increased the amount of training data. This could be because there were fewer labels to classify, and thus a smaller chance of getting it wrong, or because training was better with more data. Likely, it is a combination of both, but it is promising that with continued data collection, we will likely be able to improve classification accuracy.

Furthermore, looking at individual spectra or a few randomly selected ones per biopsy loses spatial information, which might be valuable not only for the presented classification tasks but also for the initial calculation of fluorophore abundances. Hence, these tasks could benefit from a convolutional approach considering spatial information and translational invariance. Future work will explore inputting the hyperspectral data cubes directly into a CNN. In general, we have only considered relatively simple, "off-the-shelf" models, so accuracy may be improved by considering other learning algorithms. In our tests, random forest models outperformed MLP models for every task. However, with more data or more careful design of the neural network architecture and hyperparameters, a neural network would be expected to outperform the random forests. This is left for future work as this paper focuses on establishing the feasibility of a fluorescence-based approach to tumor classification.

IDH mutations are oncogenic drivers in glioma and act as a relevant prognostic marker in glioma[67]. Thus, having this information during surgery could be valuable. In the case of eloquence, a surgeon might be less aggressive in a young patient if he/she knows that it is an IDH-mutant oligodendroglioma. In the case of an IDH-wildtype tumor, the survival chances would increase relevantly by maximizing resection and thus an aggressive surgery would be warranted. Our model can differentiate mutations in IDH with an accuracy of 93%. This is promising for future intraoperative applications.

The labeling of biopsies assumes that the entire biopsy belongs to a particular class. This is a reasonable assumption as the biopsies are small, carefully removed, and thus relatively homogeneous. In general, however, the tissue is very heterogeneous, and the distinction between classes is blurred. This could introduce noise into the margin classification training data. For example, a pixel from a biopsy labeled solid tumor might, in fact, be in a small region of infiltrating tumor and thus be incorrectly labeled. This could partially explain why the classification accuracy is only 85.7%. The only way to improve this would be to register the fluorescence image spatially with the histopathology to obtain a pixel-for-pixel map of class labels. However, the histopathological assessment is made from thin biopsy slices in unknown orientations and planes, so this is infeasible with the current setup and constitutes future work.

In addition, heterogeneity is an issue not only for an ML classifier but also for pathologists. The categorization of tumor margins can be subjective, and there is an intra-observer variability among different pathologists and centers which affects the testing dataset[68,69]. Thus, it is likely that the classification model presented in this paper would perform better if pathologists applied a standardized, quantitative measure to distinguish between different tissue regions. With this in mind, the achieved classification accuracy of 85.7% is very good. More rigorous labeling of margins could be performed in future by having several pathologists from different centers label the data.

Finally, though PCA was used in this paper primarily for visualization, analysis of information loss, and comparison to previous work[58], it should not be used as an alternative for spectral unmixing. PCA is non-unique, so the factors can be rotated to produce a set of axes that is equivalently optimal from the PCA perspective but which provides an entirely different spectral unmixing result. Instead, independent component analysis (ICA) may be more suitable, assuming the different fluorophores are statistically independent[70].

This paper has explored the effect of different neurosurgically relevant categorizations of brain tumors and tissue on the five previously-characterised[35] fluorescence emission spectra. At least four of five fluorophore abundances were found to vary statistically significantly ($p < 0.01$) among tumor margins, WHO grades, and tissue types. To test the predictive value of the unmixing, we introduced four different ML-based classifiers for tissue type, tumor margins, WHO grades and IDH status. Each classifier achieved high accuracy, thus promising practical utility for similar systems in the future and demonstrating the differential expression of the fluorophores in different tissue classes and tumors. Together with the fact that the five mathematically optimal PCA-derived components matched very closely to the physically justified fluorophores, this also shows the value and accuracy of the five fluorophores as biomarkers. Moreover, it shows potential for automatic intraoperative classification systems in fluorescence-guided neurosurgery in the future.